\documentclass[runningheads]{svmult}

\usepackage{graphicx}  
\usepackage{subeqnar}  
\usepackage{multicol}  
\usepackage{physprbb}  
\usepackage{amsmath}

\begin{document}

\title*{An Introduction to 5-Dimensional Extensions\protect\newline of the
Standard Model\footnote{Lectures given by R. R\"uckl at the International
School ``Heavy Quark Physics'', May~27~-~June~5, 2002, JINR, Dubna, 
Russia}}

\toctitle{An Introduction to 5-Dimensional Extensions
of the Standard Model}

\titlerunning{An Introduction to 5-Dimensional Extensions
of the Standard Model}

\author{Alexander M\"uck \inst{1}
\and Apostolos Pilaftsis \inst{2}
\and Reinhold R\"uckl \inst{1}
}

\authorrunning{Alexander M\"uck et al.}

\institute{Institut f\"ur Theoretische Physik und Astrophysik,
          Universit\"at W\"urzburg,\\ Am Hubland, 97074 W\"urzburg, 
          Germany
     \and Department of Physics and Astronomy, University of Manchester,\\
          Manchester M13 9PL, United Kingdom}

\maketitle              % typesets the title of the contribution

\begin{abstract}  
We give  a  pedagogical introduction  to  the  physics  of large extra
dimensions.  We  focus  our discussion on   minimal  extensions of the
Standard Model in  which  gauge fields  may   propagate in  a  single,
compact  extra   dimension while the    fermions are  restricted  to a
4-dimensional Minkowski subspace. First, the basic ideas, including an
appropriate  gauge-fixing procedure in the higher-dimensional context,
are  illustrated in simple   toy  models. Then,  we  outline  how  the
presented  techniques  can be   extended to  more realistic  theories.
Finally, we   investigate   the phenomenology  of  different   minimal
Standard Model extensions, in which all  or only some of the SU(2)$_L$
and U(1)$_Y$ gauge fields and  Higgs bosons feel  the presence of  the
fifth dimension.   Bounds on the compactification  scale between 4 and
6~TeV, depending  on the model,  are established by analyzing existing
data.
\end{abstract}

\section{Introduction}

\vspace{-12.5cm}

\begin{flushright}
MC-TH-2002-06\\
WUE-ITP-2002-025\\
hep-ph/0209371\\
\end{flushright}

\vspace{10.9cm}

Why do  we live in four  dimensions?  This fundamental  question still
cannot be answered.   However,   already  at  the beginning  of    the
20$^\mathrm{th}$ century, Kaluza and Klein realized~\cite{KK} that the
question  itself may be ill  posed.  It seems  more appropriate to ask
instead: In how many dimensions do we live?

From the  modern physics point of  view, a satisfactory  answer to the
above question may  be found within the context  of string theories or
within a more unifiable framework, known as $M$ theory.  The reason is
that  string theories  provide  the only  known theoretical  framework
within  which gravity  can  be  quantized and  so  undeniably plays  a
central r\^ole in our endeavours of unifying all fundamental forces of
nature.   A  consistent  quantum-mechanical  formulation of  a  string
theory,  however,  requires  the  existence of  additional  dimensions
beyond the four  ones we experience in our  every-day life.  These new
dimensions must  be sufficiently small, in some  appropriate sense, so
as  to  have  escaped  our  detection.   As we  will  see  in  detail,
compactification,  where additional  dimensions are  considered  to be
compact manifolds  of a characteristic size $R$,  provides a mechanism
which can  successfully hide  them.  In the  original string-theoretic
considerations~\cite{review},  the inverse length  $1/R$ of  the extra
compact dimensions and the string  mass $M_s$ turned out to be closely
tied  to  the  4-dimensional   Planck  mass  $M_{\rm  P}  =  1.9\times
10^{16}$~TeV, with all  involved mass scales being of  the same order.
More recent studies,  however, have shown~\cite{IA,JL,EW,ADS,DDG} that
there  could still be  conceivable scenarios  of stringy  nature where
$1/R$ and $M_s$ may be lowered independently of $M_{\rm P}$ by several
or many orders of magnitude.  Taking such a realization to its natural
extreme,  Ref.~\cite{ADS}  considers the  radical  scenario, in  which
$M_s$ is of order TeV and represents the only fundamental scale in the
universe at which  unification of all forces of  nature occurs.  Thus,
the  so-called  gauge hierarchy  problem  due  to  the high  disparity
between  the electroweak and  the 4-dimensional  Planck scales  can be
avoided all together, as it does not appear right from the beginning.

Let us  now try to understand  why  $n$ extra dimensions  with a large
radius $R$ can influence  gravity. This question is  tightly connected
to the geometry of space-time.    At distances small compared to  $R$,
the gravitational potential will simply change according to the  Gauss 
law in $n+4$~dimensions, i.e.
\begin{equation}
  \label{v1} 
V(r) \sim \frac{m_1 \, m_2}{M_{\mathrm{G}}^{2 + n}} \, \frac{1}{r^{n+1}} \, , 
\end{equation}
where $r\ll R$ and $M_{\mathrm{G}}$ is the true gravitational scale to
be distinguished from the Planck  scale $M_{\rm P}$.  As the distance,
at  which  gravity  is  probed,  becomes much  larger  than  $R$,  the
potential will again look effectively four dimensional, i.e.
\begin{equation} 
 \label{v2}
V(r)\ \stackrel{r\gg R}{\to}\ \frac{m_1 \, m_2}{M^2_{\mathrm{P}}} \,
\frac{1}{r}\ .
\end{equation}
Matching the two potentials (\ref{v1}) and (\ref{v2}) to give the same
answer at   $r  = R$,  we  derive  an  important  relation  among  the
parameters $M_{\rm P}$, $M_{\rm G}$ and $R$~\cite{ADS}:
\begin{equation} 
  \label{Gauss}
M^2_{\mathrm{P}} = M_{\mathrm{G}}^{2 + n} \, R^{n} \, .
\end{equation}
Hence, the weakness  of gravity, observed  by today's experiments,  is
not due to the enormity of the Planck scale $M_{\rm P}$, but thanks to
the presence of a large radius $R$.  As a result, the true fundamental
gravity scale $M_{\rm G}$  is determined from~\ref{Gauss} and is  much
smaller than $M_{\rm P}$.  For example, extra dimensions of size
\begin{equation} 
R \sim \left( \, \frac{M_{\rm P}}{M_{\rm G}} \, \right)^{\! 2/n} \,
\frac{1}{M_{\rm G}} \sim \left\{
\begin{array}{ll} 
\mathcal{O}(\mathrm{1 \, mm}), & \, \, n=2 \\ 
\mathcal{O} (\mathrm{10 \, fm}), & \, \, n=6 
\end{array} 
\right.
\end{equation}
are needed for  a gravitational scale ---typically  of the order  of a
string   scale     $M_s$---  in  the    TeV~range.     Therefore, even
Cavendish-type experiments may potentially test the model by observing
deviations  from Newton's law~\cite{ADS}  at  distances smaller than a
mm.

This low string-scale  effective model could  be embedded within e.g.\
type~I string theories~\cite{EW}, where the Standard Model (SM) may be
described as an   intersection of $Dp$   branes~\cite{ADS,DDG,AB}. The
$Dp$ brane description  implies that the SM  fields do not necessarily
feel the  presence of all the  extra dimension,  but are restricted to
some subspace of the  full space-time. Especially  mm-size dimensions,
being clearly excluded for the SM by experimental evidence, are probed
only by    gravity.    However,   as  such  intersections     may   be
higher-dimensional as  well,  in addition  to  gravitons the SM  gauge
fields could also propagate within  at least a single extra dimension.
Here, the bounds on the compactification radius from experimental data
are much more severe and $R$ has to be at least as small as an inverse
TeV.  In  our  introductory   notes,  we  will  abandon gravity    and
concentrate  on the    embedding of the    Standard  Model in  a  five
dimensional  space-time. Our main  interest  is to  explain the  basic
ideas and techniques for constructing this kind of theories.

Note  that  this limited class of  models   with low string-scales may
result   in  different   higher-di\-men\-si\-onal  extensions  of  the
SM~\cite{AB,AKT}, even if gravity is  completely ignored.  Hence,  the
actual experimental limits on the compactification radius are, to some
extent, model  dependent.      In    fact, most   of     the   derived
phenomenological limits in  the  literature were obtained  by assuming
that the   SM gauge  fields     propagate all   freely  in a    common
higher-dimensional           space~\cite{NY,WJM,CCDG,DPQ2,RW,DPQ1,CL}.
Therefore,  towards  the end  of our notes,   we will also discuss the
phenomenological  consequences of  models  which minimally depart from
the  assumption of   these  higher-dimensional   scenarios~\cite{MPR}.
Specifically, we will  consider  5-dimensional  extensions of the   SM
compactified  on an  $S^1/Z_2$  orbifold,   where the SU(2)$_L$    and
U(1)$_Y$ gauge bosons may not both live in the same higher-dimensional
space, the so-called bulk.   In all our models,   the SM fermions  are
localized on  the 4-dimensional subspace, i.e.~on a  3-brane or, as it
is often simply called, brane.

The present introductory  notes are  organized  as follows:  in  Sect.
\ref{5DQED}  we  introduce the  basic  concepts  of higher-dimensional
theories in simple  Abelian  models.   After compactifying the   extra
dimension on a  particular orbifold, $S^1/Z_2$, we obtain an effective
4-dimensional theory,  which   in addition to   the usual   SM  states
contains  infinite towers of massive  Kaluza--Klein (KK) states of the
higher-dimensional   gauge fields.  In    particular, we  consider the
question how to  consistently  quantize the higher-dimensional  models
under study  in the   so-called  $R_\xi$ gauge.  Such   a quantization
procedure can  be successfully applied to  theories  that include both
Higgs bosons  living in the bulk  and/or  on the brane.  After briefly
discussing how  these  concepts can  be  applied to  the  SM in  Sect.
\ref{MinimalExtensions}, we turn our attention to the phenomenological
aspects of  the models of our interest  in Sect.  \ref{Phenomenology}.
For each higher-dimensional   model, we calculate  the  effects of the
fifth dimension on electroweak observables and LEP2 cross sections and
analyze their    impact   on   constraining    the    compactification
scale. Technical details are omitted here in favour of introducing the
main concepts.   A  complete discussion, along  with detailed analytic
results and an  extensive list of   references, is given in  our paper
in~\cite{MPR}.  Finally, we summarize  in Sect.  \ref{Conclusions} our
main results.

\section{5-Dimensional Abelian Models}\label{5DQED}

As a starting point, let  us consider the Lagrangian of  5-dimensional
Quantum Electrodynamics (5D-QED) given by
\begin{equation}
\label{freelagrangian}
{\mathcal L} (x, y) \, = \, - \frac{1}{4} F_{M N} (x, y) F^{M N} (x, y) \:
+\: {\mathcal L}_{\mathrm{GF}}(x,y)\:,
\end{equation}
where
\begin{equation}
\label{fieldstrength}
F_{M N} (x, y) \, = \, \partial_M A_N (x, y) - \partial_N A_M (x, y)
\end{equation}
denotes  the  5-dimensional  field  strength   tensor,  and ${\mathcal
L}_{\mathrm{GF}}(x,y)$   is the  gauge-fixing  term. The Faddeev-Popov
ghost    terms  have been    neglected,    because   the ghosts    are
non-interacting in  the Abelian case.    Our notation for  the Lorentz
indices and  space-time coordinates is: $M,N  = 0,1,2,3,5$; $\mu,\nu =
0,1,2,3$; $x = (x^0,\vec{x})$; and $y = x^5$ denotes the coordinate of
the additional dimension.

The  structure  of the conventional QED  Lagrangian  is simply carried
over to the five-dimensional case. The field content  of the theory is
given by a single gauge-boson $A_M$ transforming as a vector under the
Lorentz  group SO(1,4).  In the absence  of the gauge-fixing and ghost
terms,  the   5D-QED Lagrangian    is  invariant under  a  U(1)  gauge
transformation
\begin{equation}
\label{gaugetrf}
A_M(x,y) \to A_M(x,y) + \partial_M \Theta(x,y) \, .
\end{equation}
Hence, the defining features of conventional QED are present in 5D-QED
as well.   So far, we have treated  all the spatial  dimensions on the
same footing.  This is   certainly an assumption in contradiction  not
only to experimental evidence but also to  our daily experience. There
has  to be  a  mechanism  in  the theory  which hides  the  additional
dimension  at low  energies.  As we will see    in the following,  the
simplest approach accomplishing this  goal is  compactification, i.e.,
replace the infinitely extended extra dimension by a compact object.

A simple  compact  one dimensional manifold  is  a circle,  denoted by
$S^1$, with radius $R$.  Asking for an additional  reflection symmetry
$Z_2$ with respect  to the origin  $y=0$,  one is led to  the orbifold
$S^1/Z_2$  which turns  out to  be especially  well suited for  higher
dimensional   physics.  Thus,  we    consider  the extra   dimensional
coordinate $y$ to run only from 0 to $2 \pi R$  where these two points
are identified. Moreover, according to the $Z_2$ symmetry, $y$ and $-y
= 2 \pi \! - \! y$  can be identified  in a certain sense: knowing the
field content for the segment $y \in [0,\pi]$ implies the knowledge of
the whole system. For that reason, the fixed points $y=0$ and $y=\pi$,
which do not transform under $Z_2$, are  also called boundaries of the
orbifold.

The compactification    on   $S^1/Z_2$  reflects  itself   in  certain
restrictions for the fields. In order not to  spoil the above property
of  gauge symmetry, we   demand the fields   to satisfy the  following
equalities:
\begin{equation}
\label{fieldconstraints}
\begin{split}
A_M(x,y) \, & = \, A_M(x,y + 2 \pi R)\,, \\ 
A_{\mu}(x,y) \, & = \, A_{\mu}(x, - y)\,, \\ 
A_5(x,y) \, & = \, - A_5(x, - y)\,,\\ 
\Theta(x,y) \, & = \, \Theta(x,y + 2 \pi R)\,,\\ 
\Theta(x,y) \, & = \, \Theta(x, - y)\, .
\end{split}
\end{equation}
The field $A_\mu   (x,y)$ is taken to be   even under $Z_2$, so as  to
embed conventional QED with  a massless photon  into our 5D-QED, as we
will see below.   Notice that the  reflection properties  of the field
$A_5(x,y)$ and the  gauge  parameter  $\Theta  (x,y)$ under  $Z_2$  in
(\ref{fieldconstraints})  follow automatically   if  the theory is  to
remain gauge invariant after compactification.

Making the periodicity   and  reflection properties of   $A_{\mu}$ and
$\Theta$ in~(\ref{fieldconstraints})   explicit, we can  expand  these
quantities in Fourier series
\begin{equation}
\label{fourierseries}
\begin{split}
A^{\mu}(x, y) \, & = \, \frac{1}{\sqrt{2 \pi R}} \, A^{\mu}_{(0)}(x)
                     \, + \, \sum_{n=1}^{\infty} \, \frac{1}{\sqrt{\pi
                     R}} \, A^{\mu}_{(n)}(x) \, \cos \left( \,  \frac{n
                     y}{R} \, \right)  \, , \\
\Theta (x, y) \, & = \, \frac{1}{\sqrt{2 \pi R}}
                     \, \Theta_{(0)}(x) \, + \, \sum_{n=1}^{\infty} \,
                     \frac{1}{\sqrt{\pi R}} \, \Theta_{(n)}(x) \cos
                     \left( \frac{n y}{R} \right) \, .
\end{split}
\end{equation}
The    Fourier   coefficients   $A_{(n)}^{\mu}(x)$  are the  so-called 
Kaluza-Klein (KK) modes. The   extra component  of   the  gauge  field 
is  odd  under the reflection symmetry and its expansion is given by
\begin{equation}
A^5(x, y) \, = \, \sum_{n=1}^{\infty} \, \frac{1}{\sqrt{\pi R}} \,
                     A^5_{(n)}(x) \, \sin \left( \,  \frac{n y}{R} \, \right) \, . 
\end{equation}
Note  that there is no zero mode, a phenomenologically important fact, 
as we will see below.

At this  point, the  theory is again  formulated entirely in  terms of
four-di\-men\-sio\-nal fields, the KK modes. All the dependence of the
Lagrangian density  on the extra coordinate $y$  is parameterized with
simple  Fourier functions.  Finally,  the physics  is dictated  by the
Lagrangian  anyway, not  by its  density, thus,  one can  go  one step
further  and completely  remove  the explicit  $y$  dependence of  the
Lagrangian by integrating  out the extra dimension.  From  now on, the
quantity of interest will be
\begin{equation}
\label{integration}
\mathcal{L}(x) \, = \, \int_0^{2 \pi R} \, dy \, \, \mathcal{L}(x, y) \, .
\end{equation}
All the higher-dimensional physics is  reflected by the infinite tower
of KK modes for each field component. A  simple calculation yields the
4-dimensional Lagrangian
\begin{equation}
\label{labeforegf}
\begin{split}
{\mathcal L}(x) =& -\, \frac{1}{4} F_{(0) \mu \nu} \, F_{(0)}^{\mu \nu} + 
   \sum_{n = 1}^{\infty} \, \left[ \, -\frac{1}{4} \, F_{(n) \mu \nu} \,
   F_{(n)}^{\mu \nu} \right.\\
& \left. +\, \frac{1}{2} \left( \, \frac{n}{R} \, A_{(n) \mu} +
   \partial_{\mu} A_{(n) 5} \, \right) \left( \, \frac{n}{R} \, A^\mu_{(n)} +
   \partial^\mu A_{(n)5} \, \right) \, \right]\: +\:  {\mathcal L}_{\mathrm{GF}}(x) \, ,
\end{split}
\end{equation}
where ${\mathcal   L}_{\mathrm{GF}}(x)$   is  defined in   analogy  to
(\ref{integration}). The  first term in  (\ref{labeforegf}) represents
conventional  QED involving the  massless field  $A^{\mu}_{(0)}$. Note
that all   the other  vector   excitations  $A^{\mu}_{(n)}$ from   the
infinite tower of KK  modes come with  mass terms, their mass being an
integer multiple of the  inverse compactification radius. Therefore, a
small radius  leads to a  large mass or compactification scale $M=1/R$
in the model. It is this large scale which is responsible for the fact
that  an extra   dimension, as  it   may   exist,  has not   yet  been
discovered.    The extra dimension  is,  so  to speak,   hidden by its
compactness.

Note that  it is the absence  of $A^{5}_{(0)}$ due   to the odd  $Z_2$
symmetry of  $A^{5} (x,y)$ which allows us to recover conventional QED 
in the low energy limit of the model. For  $n   \ge 1$,  the KK  tower
$A^{5}_{(n)}$ for the additional  component  of  the five  dimensional
vector field  mixes  with  the vector modes. The  modes $A^{5}_{(n)}$,
being scalars  with  respect to the   four  dimensional Lorentz group,
play the  r\^ole of  the would-be  Goldstone  modes  in  a  non-linear
realization  of an Abelian  Higgs  model, in which  the  corresponding
Higgs fields are taken to be infinitely  massive. Thus, one is tempted
to  view   the  mass    generation   for  the     heavy KK   modes  by
compactification  as  a   kind  of geometric   Higgs  mechanism. Note,
moreover, that  the  Lagrangian (\ref{labeforegf}) is still manifestly
gauge  invariant  under  the transformation (\ref{gaugetrf})  which in
terms of the KK modes reads
\begin{equation}
\label{gaugetrfbefresc}
\begin{split}
A_{(n) \mu}(x) \, & \to \, A_{(n) \mu}(x) + \partial_{\mu} \Theta_{(n)}(x)  \, , \\
A_{(n) 5}(x) \, & \to \, A_{(n) 5}(x) - \frac{n}{R} \Theta_{(n)}(x) \, .
\end{split}
\end{equation}

The above observations  motivate us to  seek for a  higher-dimensional
generalization of 't-Hooft's gauge-fixing   condition, for which   the
mixing   terms  bilinear in   $A^{\mu}_{(n)}$   and  $A^{5}_{(n)}$ are
eliminated        from     the      effective            4-dimensional
Lagrangian~(\ref{labeforegf}).  Taking  advantage  of  the fact   that
orbifold     compactification      generally      breaks       SO(1,4)
invariance~\cite{GGH}, one can abandon  the requirement  of covariance
of the gauge fixing condition with  respect to the extra dimension and
choose    the  following   non-covariant   generalized   $R_\xi$ gauge
\cite{MPR,GNN}:
\begin{equation}
\label{gengaugefixterm}
{\mathcal L}_{\mathrm{GF}}(x, y)\ =\ -\, \frac{1}{2 \xi} (\partial^{\mu} A_{\mu}
\: - \: \xi \, \partial_5 A_5)^2 \, .
\end{equation}
Nevertheless,  the  gauge-fixing  term  in (\ref{gengaugefixterm})  is
still    invariant      under    ordinary    4-dimen\-sional   Lorentz
transformations. Upon integration over the extra dimension, all mixing
terms   in  (\ref{labeforegf})   drop   out  up  to  irrelevant  total
derivatives. Thus, the gauge-fixed  four dimensional Lagrangian of the
5-dimensional QED explicitly shows the different degrees of freedom in
the model. It reads
\begin{equation}
\label{lagrangedensity}
\begin{split}
\mathcal{L}(x) \, & = \, 
- \frac{1}{4} F_{(0) \mu \nu} \,  F_{(0)}^{\mu \nu} \, - \, \frac{1}{2 \xi} \, 
(\partial^{\mu} A_{(0) \mu})^2\\
& \quad \quad + \sum_{n = 1}^{\infty} \, \left[ -\frac{1}{4} \, F_{(n) \mu \nu} \,  
F_{(n)}^{\mu \nu} \, + \frac{1}{2} \, \left( \, \frac{n}{R} \, \right)^2 \, 
A_{(n)}^{\mu} \, A_{(n)\mu} - \, \frac{1}{2 \xi} \, (\partial^{\mu} 
A_{(n) \mu})^2 \right]\\
& \quad \quad + \sum_{n = 1}^{\infty} \, \left[ \frac{1}{2} \, 
(\partial^{\mu} A_{(n) 5}) \, (\partial_{\mu} A_{(n) 5}) \, - \,  
\frac{1}{2} \, \xi \, \left( \, \frac{n}{R} \, \right)^2 \, A_{(n) 5} \, ^2 \right] \, .
\end{split}
\end{equation}
Gauge fixed QED is accompanied with a tower of its massive copies. The
scalars $A_{(n) 5}$ with gauge  dependent masses resemble the would-be
Goldstone bosons of  an ordinary 4-dimensional Abelian-Higgs model  in
the $R_\xi$  gauge.   From this  Lagrangian, it  is  obvious that  the
corresponding propagators take on their usual forms:
\begin{equation}
\label{propagators}
\includegraphics[width=10cm]{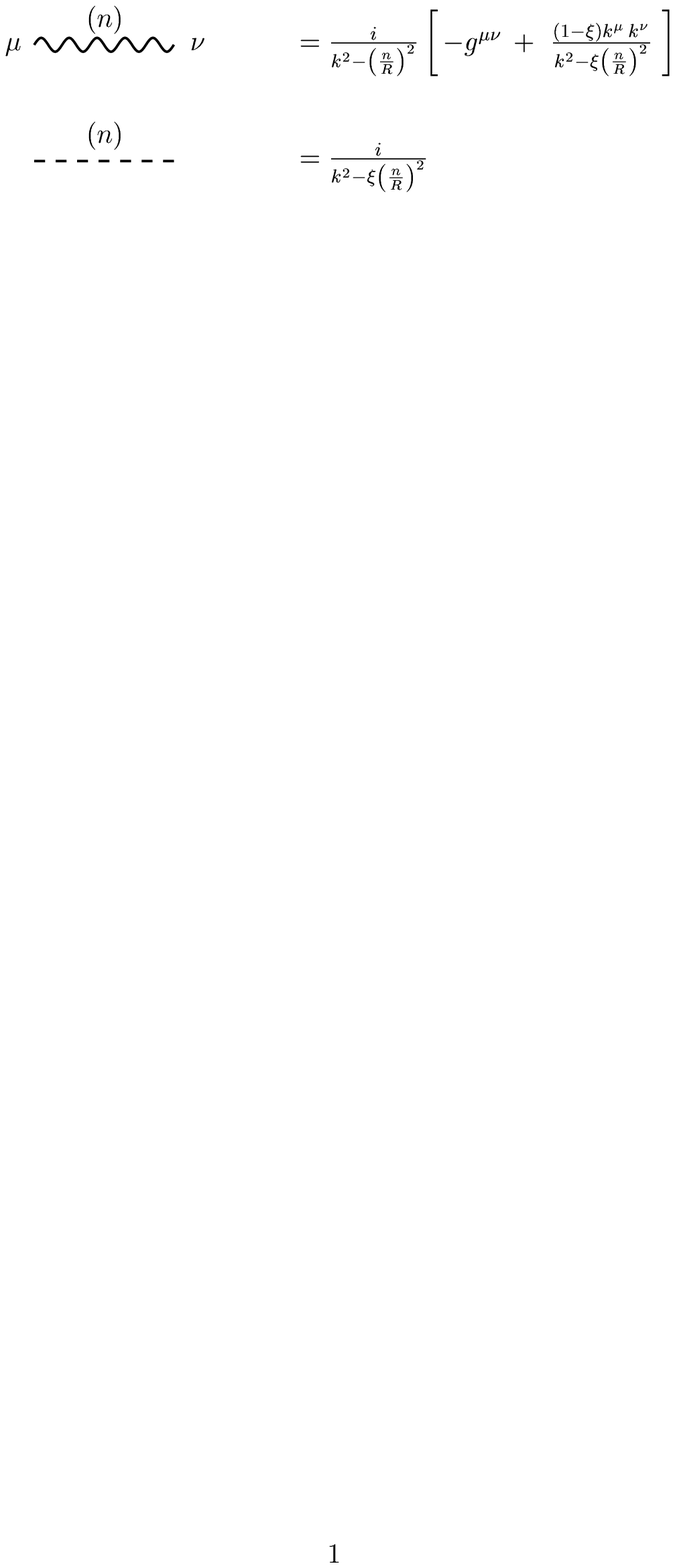}
\end{equation}
Hereafter,  we shall refer  to the  $A^{5}_{(n)}$ fields  as Goldstone
modes.

Having defined the appropriate  $R_\xi$ gauge through the gauge-fixing
term   in~(\ref{gengaugefixterm}), we  can recover   the usual unitary
gauge, in which  the Goldstone modes  decouple from the theory, in the
limit $\xi \to \infty$ \cite{PS,DMN}.  Thus,  for the case at hand, we
have  seen  how    starting from a   non-covariant  higher-dimensional
gauge-fixing  condition,  we can  arrive     at the  known   covariant
4-dimensional $R_\xi$ gauge after compactification.

\begin{figure}[t]
\begin{center}
\includegraphics[width=9cm]{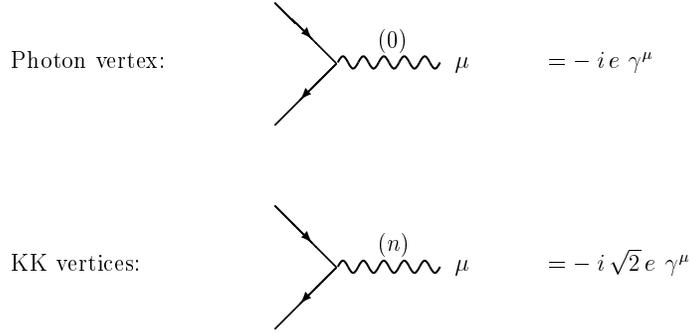}
\end{center}
\caption{\label{QEDvertices} Feynman rules for the vertices in 5D-QED}
\end{figure}

Having  established  a  five  dimensional  gauge sector,  we  can  now
introduce fermions in  the model. This is possible  in the same spirit
followed for the gauge field,  leading to bulk fermions, i.e. fermions
propagating into  the extra dimension~\cite{IA,DPQ2,ACD}.  However, it
turns  out,  that  there  is  an even  easier  and  phenomenologically
challenging  alternative to  this approach.   Moreover,  problems with
chiral fermions in  five dimensions, to be included  in more realistic
theories, can be avoided.  The $S^1/Z_2$ orbifold, as noted above, has
the peculiar feature that there are fixed points $y = 0$ and $y = \pi$
not transforming  under the $Z_2$ symmetry.  These  special points can
be  considered  as so-called  branes  hosting  localized fields  which
cannot  penetrate the  extra  dimension. This  concept  can be  easily
formalized  by  introducing  a  $\delta$-function  in  the  Lagrangian
for the fermions, i.e.
\begin{equation}
\label{fermladens}
\mathcal{L}_{\mathrm{F}}(x,y) \, = \, \delta(y) \, \overline{\Psi}(x) \, 
\left( i \, \gamma^{\mu} \, D_{\mu} \, - \, m_f \right) \, \Psi(x) \, ,
\end{equation}
where the covariant derivative 
\begin{equation}
\label{covderin5DQED}
D_{\mu} \, = \, \partial_{\mu} \, + \, i \, e_5 \, A_{\mu}(x,y) 
\end{equation}
contains the bulk gauge field and  $e_5$ denotes the coupling constant
of   5D-QED.    The     obvious   generalization    for  the     usual
gauge-transformation properties of fermion fields reads
\begin{equation}
\label{gaugetrfferm5DQED}
\Psi(x) \to \exp \left( -i \, e_5 \, \Theta(x,0) \right) \Psi (x) \, .
\end{equation}
Again integrating   out the  fifth dimension,   we  are  left  with an
effective four dimensional interaction Lagrangian
\begin{equation}
\label{interactionsfermwithgbmodesabelianmodel}
\mathcal{L}_{\mathrm{int}}(x)\ = \ - \, e \, \overline{\Psi}\, \gamma^{\mu} \, \Psi\, 
			\left( \, A_{(0)\mu} \, + \, \sqrt{2} \, \sum_{n=1}^{\infty}
                        A_{(n) \mu} \, \right) \, ,
\end{equation}
coupling  all the KK  modes to  the fermion field   on the brane.  The
coupling   constant $e  =  e_5/\sqrt{2 \pi  R}$   is the QED  coupling
constant  as  measured   by  experiment.  The    factor  $\sqrt{2}$ in
(\ref{interactionsfermwithgbmodesabelianmodel})     is     a   typical
enhancement factor for   the coupling  of  brane  fields to heavy   KK
modes. Note that the scaler modes $A^5_{(n)}$ do  not couple at all to
brane fermions because their wave functions vanish at $y=0$ according
to the odd   $Z_2$-symmetry.  These interaction terms   together  with
completely standard  kinetic  terms  for  the fermion field   complete
5D-QED. The corresponding Feynman rules for the electron-photon vertex
and  the  analogous   interaction of   the    KK modes are    shown in
Fig. \ref{QEDvertices}.

\begin{figure}[t]
\begin{center}
\includegraphics[width=9cm]{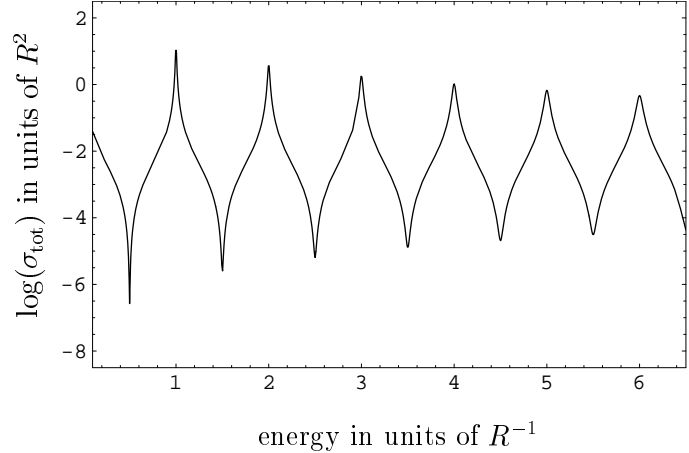}
\end{center}
\caption{\label{crosssection} Total cross section for $e^+ e^- \to \mu^+ \mu^-$ as a 
function of center-of-mass energy on a logarithmic scale. The width of the KK modes has
been reasonably approximated  }
\end{figure}

If nature  were  described by QED  up  to energies  probed so   far by
experiment    an  experimental   signature of  this   five dimensional
extension would  be,   e.g., a  series  of $s$-channel  resonances  in
muon-pair  production  at  an  $e^+   e^-$-collider as  shown  in Fig.
\ref{crosssection}.  Even though nature is not  described by QED only,
the generic signatures of extra dimensions are  quite similar to those
in more realistic theories.

The above quantization procedure can now be extended to more elaborate
higher-dimen\-sional models.  If we want to  extend the Standard Model
by an  extra dimension  we   have to understand   spontaneous symmetry
breaking in this context.  Hence, adding a  Higgs scalar in the  bulk,
the 5D Lagrangian of the theory reads
\begin{equation}
\label{lagrdens5Dabelianhiggsmodel}
{\mathcal L} (x,y) \ = \ - \, \frac{1}{4} \, F^{M N}\, F_{M N}\: +\:
               (D_M\Phi)^*\, (D^M \Phi ) \: - \: V(\Phi )
               \: + \: {\mathcal L}_{\mathrm{GF}}(x,y) \, ,
\end{equation}
where  $D_M$     again      denotes    the  covariant       derivative
(\ref{covderin5DQED}), $e_5$ the 5-dimensional gauge coupling,
\begin{equation}
\Phi  (x,y)\ = \ \frac{1}{\sqrt{2}} \, 
(\, h(x,y)  \, +  \, i  \, \chi(x,y)\, )
\end{equation}
a 5-dimensional complex scalar field, and 
\begin{equation}
V(\Phi) \, =  \, \mu_5^2 \, | \Phi |^2 \, + \,
\lambda_5 \, | \Phi |^4\, 
\end{equation}
(with $\lambda_5 > 0$) the 5-dimensional Higgs potential.  We consider
$\Phi  (x,y)$ to be even under  $Z_2$, perform a corresponding Fourier
decomposition, and integrate over $y$ to obtain
\begin{equation}
\label{kintermsforhiggsinabmodel}
\mathcal{L}_{\mathrm{H}} (x) \! = \! \frac{1}{2} \, \sum_{n=0}^{\infty}\left[ \,
(\partial_{\mu} h_{(n)})\,(\partial^{\mu} h_{(n)})\: -\:
\frac{n^2}{R^2}\, h_{(n)}^2\: - \: 
\mu^2 \, h_{(n)}^2\: + \: \left( \, h \leftrightarrow \chi \, \right) \right] 
\! +\! \, \dots \,,
\end{equation}
For $\mu^2 =  \mu^2_5 < 0$, as  in  the usual  4-dimensional case, the
zero KK Higgs mode  acquires a non-vanishing vacuum  expectation value
(VEV) which breaks the U(1) symmetry.   Moreover, it can be shown that
as long as the phenomenologically relevant condition $v < 1/R$ is met,
$h_{(0)}$ will be the only mode to receive a non-zero VEV
\begin{equation}
\langle h_{(0)} \rangle = v = \sqrt{2 \pi R \, |\mu_5|^2/ \lambda_5} \, .
\end{equation}

The VEV  introduces an additional  mass term for  each  KK mode of the
gauge fields. The zero mode turns from a massless  to a massive degree
of freedom, as usual for the Higgs mechanism. All the higher KK masses
are  slightly  shifted. The gauge  and  self interactions of the Higgs
fields,  omitted  in  (\ref{kintermsforhiggsinabmodel}), only  involve
bulk fields, in contrast to  the photon-fermion interaction introduced
before. Although  this leads to interesting  effects we postpone their
discussion  until Sect.  \ref{MinimalExtensions} where  we investigate
the          phenomenologically      more interesting      gauge-boson
self-couplings. After spontaneous symmetry breaking, it is instructive
to introduce the fields
\begin{equation}
\label{physunphysfieldshiggsinbulk}
G_{(n)} \, = \, \Big(\, \frac{n^2}{R^2}\: +\: e^2v^2\,
\Big)^{-1/2}\, \left( \, \frac{n}{R} \, A_{(n) 5} \: 
+\: ev\,\chi_{(n)} \, \right) \, , 
\end{equation}
where again  $e  = e_5 / \sqrt{2  \pi  R}$, and the  orthogonal linear
combinations $a_{(n)}$.   In the effective  kinetic Lagrangian  of the
theory for the $n$-KK mode ($n>0$)
\begin{equation}
\label{Lkineff}
\begin{split}
\mathcal{L}^{(n)}_{\mathrm{kin}} (x)\, = & \, 
-\, \frac{1}{4} \, F_{(n)}^{\mu \nu} \, F_{(n) \mu \nu} \\
& \, \, + \, \frac{1}{2}\, \left( m_{A (n)} \, A_{(n) \mu} \, + \, 
                              \partial_{\mu} \, G_{(n)} \right)\,
\left( m_{A (n)} \, A^\mu_{(n)} \, + \, 
                              \partial^{\mu} \, G_{(n)} \right)  \\[1ex]
   & \, \, + \, \frac{1}{2}\, (\partial_\mu a_{(n)}) \,
(\partial^\mu a_{(n)})\: - \: 
\frac{1}{2} m^2_{a (n)} a_{(n)}^2 \ +\ \dots\, ,
\end{split}
\end{equation}
$G_{(n)}$ now plays the r\^ole of a Goldstone mode in an Abelian Higgs
model.  Both,  $A_{(n)  5}$ and $\chi_{(n)}$   take  part in the  mass
generation for the heavy KK modes  and, therefore, they are also mixed
in the corresponding  Goldstone  mode. Because the  mass  contribution
from spontaneous symmetry breaking is expected to be small compared to 
the  KK  masses,  the  Goldstone  modes  are  dominated  by  the extra
component  of  the  gauge  field.  The  pseudoscalar  field  $a_{(n)}$ 
describes an additional physical KK excitation degenerate in mass with
the KK gauge mode~$A_{(n) \mu}$, i.e.
\begin{equation}
m^2_{a (n)} = m^2_{A (n)} = (n^2/R^2) + e^2v^2 \, . 
\end{equation}
The spectrum of  the zero KK modes  is simply identical  to that of  a
conventional  Abelian  Higgs  model, as  it   should be if  we  are to
rediscover known physics  in the low energy  limit.  It becomes  clear
that     the         appropriate    gauge-fixing   Lagrangian       in
(\ref{lagrdens5Dabelianhiggsmodel})  for a  5-dimensional  generalized
$R_{\xi}$-gauge should be
\begin{equation}
\label{gaugefixingfunctionAbelianmodel}
{\mathcal L}_{\mathrm{GF}} (x,y) \, = \, - \, \frac{1}{2 \xi} \, \bigg[\,\partial_{\mu}
                         A^{\mu} \, -  \, \xi \, \bigg(\,\partial_5 A_5
           \: + \: e_5 \frac{v}{\sqrt{2 \pi R}} \, \chi\,\bigg)\,\bigg]^2 \, .
\end{equation}
All the mixing  terms are removed and we  again arrive at the standard
kinetic Lagrangian  for  massive gauge  bosons and   the corresponding
would-be Goldstone modes
\begin{equation}
\label{gaugefixedlaabbulk}
\begin{split}
\mathcal{L}^{(n)}_{\mathrm{kin}} (x)\ =\ & 
- \frac{1}{4} \, F_{(n)}^{\mu \nu} \, F_{(n) \mu \nu} 
        \: + \: \frac{1}{2} \, m_{A (n)}^2 \, A_{(n) \mu}\,A^\mu_{(n)}
\: - \: \frac{1}{2 \xi} (\partial_{\mu} \, A_{(n)}^{\mu})^2 \\[1ex]
& + \, \frac{1}{2} 
(\partial_{\mu} G_{(n)})\,(\partial^{\mu} G_{(n)})\:
- \: \frac{\xi}{2} \, m_{A (n)}^2 \, G_{(n)}^2\\[1ex]
    & + \, \frac{1}{2} (\partial_{\mu} a_{(n)})\,(\partial^{\mu} a_{(n)})
\: - \: \frac{1}{2} \, m_{a (n)}^2 \, a_{(n)}^2 \\[1ex]
&+\, \frac{1}{2} (\partial_{\mu} h_{(n)})\,(\partial^{\mu} h_{(n)})\: 
- \: \frac{1}{2}\, m_{h (n)}^2 \, h_{(n)}^2 \, \, \, .
\end{split}
\end{equation}
The CP-odd scalar   modes $a_{(n)}$ and  the Higgs  KK-modes $h_{(n)}$
with mass
\begin{equation}
m_{h (n)}  =  \sqrt{(n^2/R^2) + \lambda_5  v^2  /  \pi  R}  
\end{equation}
are not affected by the gauge fixing  procedure.  Observe finally that
the  limit $\xi \to  \infty$  consistently corresponds to the  unitary
gauge.

As a qualitatively different way of implementing the Higgs sector in a
higher-dimensional  Abelian model, we can  localize the Higgs field at
the $y=0$ boundary of the $S^1/Z_2$ orbifold, following the example of
the fermions  in 5D-QED. Introducing the appropriate $\delta$-function
in the 5-dimensional Lagrangian, this amounts to
\begin{equation}
\label{lagrdens5Dabelianhiggsmodelbrane}
{\mathcal L} (x,y)\ =\ - \, \frac{1}{4} \, F^{M N}\, F_{M N}\: +\: 
\delta (y) \, \left[ \, (D_{\mu} \Phi)^* \, (D^\mu \Phi)  \: 
                - \: V(\Phi ) \, \right] \: +\:  {\mathcal L}_{\mathrm{GF}}(x,y) \, ,
\end{equation}
where the  covariant  derivative is given by (\ref{covderin5DQED}) and 
the  Higgs potential has its familiar 4-dimensional form.  Because the
Higgs potential is effectively four dimensional the  Higgs  field, not
having KK excitations as a brane field, acquires the usual VEV. Notice 
that  the  bulk  scalar  field  $A_5  (x,y)$, as  a  result of its odd 
$Z_2$-parity, does not couple to the Higgs sector on a brane.

After    compactification  and  integration   over the  $y$-dimension,
spontaneous symmetry breaking  again generates masses  for all the  KK
gauge modes $A^{\mu}_{(n)}$.  However, the mass  matrix for the simple
Fourier  modes in (\ref{fourierseries})  is no longer diagonal because
of               the             $\delta$-function                  in
(\ref{lagrdens5Dabelianhiggsmodelbrane}). Instead, it is given by
\begin{equation}
\label{massmatrixvectorbosonshiggsonbrane}
M_{A}^2 \, = \, 
\left(
\begin{array}{cccc}
m^2 \, \, \, & \sqrt{2} \, m^2 \, \, \, & \sqrt{2} \, m^2 & \cdots \\
\sqrt{2} \, m^2 \, \, \, & 2 \, m^2 + (1/R)^2 \, \, \, & 2 \, m^2 & \cdots \\
\sqrt{2} \, m^2 \, \, \, & 2 \, m^2 \, \, \, & 2 \, m^2 + (2/R)^2 & \cdots \\
\vdots \, \, \, & \vdots \, \, \, & \vdots & \ddots
\end{array}
\right) \quad ,
\end{equation}
where  $m =  ev$.  Therefore, the  Fourier  modes  are no  longer mass
eigenstates. By  diagonalization   of   the  mass  matrix    the  mass
eigenvalues $m_{(n)}$ of the KK mass eigenstates are found to obey the
transcendental equation
\begin{equation}
\label{transcendentalequformasses}
m_{(n)} \, = \, \pi \, m^2 \, R \, \cot \left( \, \pi
\, m_{(n)} \, R \, \right) \, .
\end{equation}
Hence, the zero-mode mass  eigenvalues are slightly shifted  from what
we expect in a 4D model. An approximate calculation, to first order in
$m^2/M^2$, yields
\begin{equation}
\label{massofgroundmode}
m_{(0)} \, \approx \, \left( \, 1 - 
\frac{\pi^2}{6} \,  \frac{m^2}{M^2} \, \right) \, \, m \, .
\end{equation}
The   respective   KK mass    eigenstates  can   also  be   calculated
analytically. They are given by
\begin{equation}
\label{masseigenstatesexplicit}
\hat{A}^{\mu}_{(n)} \! = \! \left( 1 + \pi^2 \, m^2 \, R^2 +
\frac{m^2_{(n)}}{m^2}\right)^{\! \! \! - 1/2} \sum^{\infty}_{j=0} \,
\frac{2 \, m_{(n)} \, m}{m^2_{(n)} \, - \, (j / R)^2} \,
\left(\frac{1}{\sqrt{2}}\right)^{\delta_{j,0}} \, A^{\mu}_{(j)} \, .
\end{equation}
The   couplings  of  these  mass  eigenstates  to  fermions  will   be
slightly shifted with respect to the couplings of the Fourier modes in
(\ref{interactionsfermwithgbmodesabelianmodel}).  To be specific,  the
interaction Lagrangian can be parameterized by
\begin{equation}
\label{couplingtocurrentonbraneintermsofmasseigenstates}
\mathcal{L}_{\mathrm{int}} \, = \, - \, \overline{\Psi} \, \gamma_{\mu} \Psi \, 
                 \sum_{n=0}^{\infty} \, e_{(n)} \, \hat{A}^{\mu}_{(n)} \, ,
\end{equation}
where the  couplings $e_{(n)}$ of the  different  mass eigenstates are
given by
\begin{equation}
\label{redefinedcouplingsabhiggsonbrane}
e_{(n)} \, = \, \sqrt{2} \, e \, \left( \, 1 \, + \, \frac{m^2}{m^2_{(n)}} \, 
+ \, \pi^2 \, \frac{m^2}{M^2} \frac{m^2}{m^2_{(n)}} \, \right)^{- \frac{1}{2}} \quad .
\end{equation}
For  example,  the shift in  the  zero  mode coupling is approximately
given by
\begin{equation}
\label{zeromasseigenstatecouplingabelianmodel}
e_{(0)} \, \approx \, 
\left( \,1 \, - \, \frac{\pi^2}{3} \, \frac{m^2}{M^2} \, \right) e \, .
\end{equation}
Here, in the Abelian  model, the shifts   in masses and couplings  may
seem to be  a mere matter  of redefinition of  the measured masses and
coupling in    terms   of   the    fundamental  constants   of     the
5D-theory.  However,   they   lead  to    important   phenomenological
implications in the context of  the higher-dimensional Standard Model,
where the various couplings are  affected differently, as we will  see
below.

To find the  appropriate  form of   the gauge-fixing  term  ${\mathcal
L}_{\mathrm{GF}}(x,y)$ in~(\ref{lagrdens5Dabelianhiggsmodelbrane}), we
follow (\ref{gaugefixingfunctionAbelianmodel}),  but     restrict  the
scalar field $\chi$ to the brane $y = 0$, viz.
\begin{equation}
\label{gaugefixingfunctionabelianhiggsonbrane}
{\mathcal L}_{\mathrm{GF}} (x,y) \, = \, - \, \frac{1}{2 \xi} \, 
        \Big[\, \partial_{\mu} A^{\mu} \ - \ 
\xi \, \big( \partial_5 \, A_5 \: + \: 
e_5 v \, \chi \, \delta(y)\big)\,\Big]^2 \, .
\end{equation}
As  is expected from a generalized  $R_\xi$ gauge, all mixing terms of
the gauge modes $A_{(n)}^{\mu}$ with $A_{(n)  5}$ and $\chi$ disappear
up to total derivatives  if $\delta (0)$ is appropriately  interpreted
on    $S^1/Z_2$.  Determining the   unphysical   mass spectrum of  the
Goldstone modes, we find a one-to-one  correspondence of each physical
vector mode of  mass $m_{(n)}$ to  an  unphysical Goldstone mode  with
gauge-dependent mass $\sqrt{\xi}  \, m_{(n)}$.   In the unitary  gauge
$\xi   \to\infty$, the would-be Goldstone   modes  are absent from the
theory.  The present  brane-Higgs   model does not  predict  other  KK
massive scalars apart from the physical Higgs boson $h$.

At  this  point, we  cannot  decide by  any   means which of   the two
possibilities for the Higgs sector,  brane or bulk Higgs fields, could
be  realized in nature.  Thus, we have to be  ready to analyze both of
them phenomenologically when we move on to SM extensions.

\section{5-Dimensional Extensions of the Standard Model}
\label{MinimalExtensions}

It is a straightforward exercise to generalize the ideas introduced in
Sect. \ref{5DQED} for non-Abelian theories
\begin{equation}
\label{nonAbelianlagrangian}
\begin{split}
\mathcal{L}(x, y) \, = \, & - \, \frac{1}{4} \, F^a_{M N} F^{a M N} 
+ \delta(y) \, \overline{\Psi}(x) \, \left( \, i \gamma^{\mu} \, D_{\mu} \, 
- \, m_f \right) \, \Psi(x)\\[1ex]
& \quad \quad + \, \mathcal{L}_{\mathrm{H}} \, + \, \mathcal{L}_{\mathrm{GF}} \, 
+ \, \mathcal{L}_{\mathrm{FP}} \, ,
\end{split}
\end{equation}
where the field  strength for the non-Abelian gauge  field of  a group
with  structure constants  $f^{abc}$  and coupling  constant $g_5$  is
given by
\begin{equation}
\label{nonAbelianfieldstrength}
F^a_{M N} \, = \, \partial_{M}A^a_{N} \, - \, \partial_{N}A^a_{M} \, 
+ \, g_5 f^{abc} A^b_{M} A^c_{N} \, .
\end{equation}
Compactification,   spontaneous  symmetry breaking   and gauge  fixing
\cite{MPR,DCH}  are   very  analogous  to the  Abelian    case and the
non-decoupling ghost sector can be easily included~\cite{MPR}.  Hence,
in the effective  4D theory, we arrive   at a particle spectrum  being
similar to the Abelian case.

\begin{figure}[fp]
\begin{center}
\includegraphics[width=12.2cm]{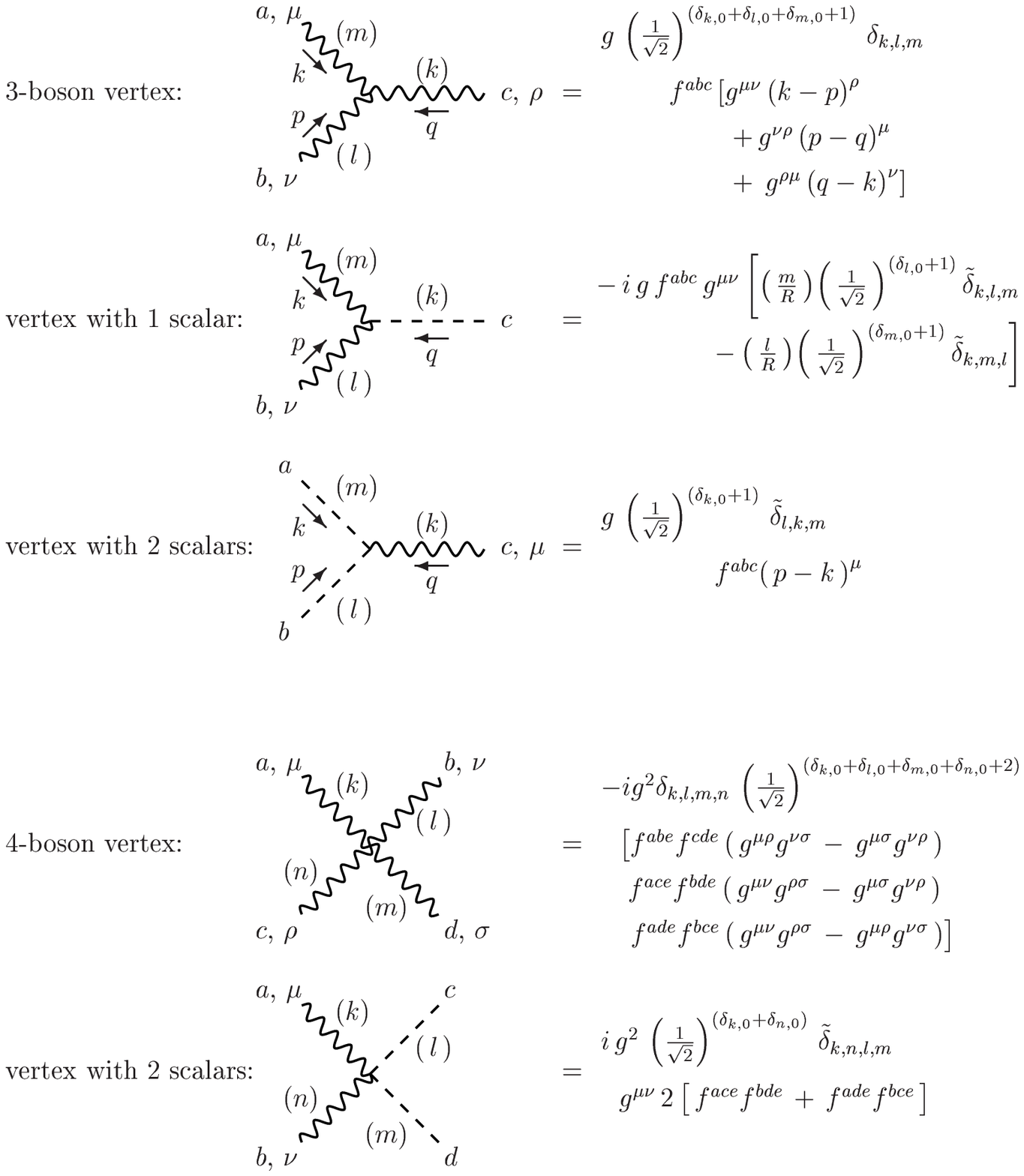}
\end{center}
\caption{\label{feynrulesnonabelian} 
Feynman  rules for the triple and  quartic gauge boson couplings.  The
scalar   modes  correspond     to   the  extra      component  of  the
higher-dimensional  gauge fields. As  in the  Abelian  model, they are
would-be  Goldstone modes and can be  gauged away in unitary gauge. In
parenthesis,  the  KK number of the   interacting modes is shown.  The
selection rules  are    enforced   by  the  prefactors    defined   in
(\ref{selectionrules}) }
\end{figure}

In addition, the self-interactions of the gauge bosons, induced by the
bilinear     terms      in    the    non-Abelian    field     strength
(\ref{nonAbelianfieldstrength}), lead  to  self-interactions of the KK
modes which are restricted by selection rules, i.e., there are certain
conditions for the KK numbers to be obeyed  at each triple and quartic
gauge-boson  vertex. This is  a general  feature  for  interactions in
which  only bulk  modes take   part.  A brane   completely breaks  the
translational invariance of the orbifold  and, thus, a brane field can
couple to any bulk mode. In contrast, interactions between bulk fields
obey a kind  of quasi-momentum conservation  with respect to the extra
dimension, reflecting the special  structure of the $S^1/Z_2$ orbifold
and leading to the  selection rules.  The corresponding  Feynman rules
are displayed in  Fig. \ref{feynrulesnonabelian}.  The selection rules
are enforced by the prefactors
\begin{equation}
\label{selectionrules}
\begin{split}
\delta_{k,l,m} & = \delta_{k+l+m,0} + \delta_{k+l-m,0} + 
\delta_{k-l+m,0} + \delta_{k-l-m,0} \, , \\
\tilde{\delta}_{k,l,m} & = - \delta_{k+l+m,0} + \delta_{k+l-m,0} 
- \delta_{k-l+m,0} + \delta_{k-l-m,0} \, , 
\end{split}
\end{equation}
where        $\delta_{i,j}$    denotes       a   standard    Kronecker
symbol. $\delta_{k,l,m,n}$ and $\tilde{\delta}_{k,l,m,n}$ are  defined
analogously.

At this point, we have considered all the important generic aspects of
higher-dimensional theories. Therefore,  we can now turn our attention
to the  theory we are really interested  in, the electroweak sector of
the Standard  Model. Its gauge   structure SU(2)$_L \otimes  $U(1)$_Y$
opens up several  possibilities for 5-dimensional extensions,  because
the SU(2)$_L$  and  U(1)$_Y$  gauge  fields  do not   necessarily both
propagate  in the extra dimension. As  the  fermion or Higgs fields we
encountered before, one of the gauge groups can be confined to a brane
at $y = 0$. Such  a realization of  a higher-dimensional model may  be
encountered within specific stringy frameworks~\cite{AB,AKT}.

However, in  the most frequently investigated  scenario, SU(2)$_L$ and
U(1)$_Y$  gauge  fields  live  in  the bulk  of  the  extra  dimension
(bulk-bulk  model).   The  Lagrangian  is  simply  an  application  of
(\ref{nonAbelianlagrangian}) to the Standard Model gauge groups. Here,
it is  possible, even before  integrating out the extra  dimension, to
choose a basis  for the fields, where the  photon and $Z$-boson fields
become  explicit.   The photon  sector  resembles  exactly the  5D-QED
discussed in  Sect.  \ref{5DQED}, while for  the $Z$ boson  and its KK
modes  spontaneous  symmetry  breaking   leads  to  the  effects  also
presented  in  Sect. \ref{5DQED}.   In  the  bulk-bulk  model, both  a
localized  (brane) and  a 5-dimensional  (bulk) Higgs  doublet  can be
included in the theory.  For  generality, we will consider a 2-doublet
Higgs  model, where  the one  Higgs field  $\Phi_1$ propagates  in the
fifth  dimension, while  the  other one  $\Phi_2$  is localized.   The
phenomenology presented in these notes  is not sensitive to details of
the Higgs potential but only  to their vacuum expectation values $v_1$
and $v_2$,  or equivalently  to $\tan \beta  = v_2  / v_1$ and  $v^2 =
v^2_1 + v^2_2$.  Hence, $\beta$  is the only additional free parameter
introduced in the model.

The chiral  structure of the Standard Model can be easily incorporated
as long as one only considers fermions restricted to a brane. A simple
extension  of (\ref{interactionsfermwithgbmodesabelianmodel}) leads to
\begin{equation}
\label{genericfermionint}
{\mathcal L}_{\mathrm{int}}(x)\ = \ g \, \overline{\Psi} \, \gamma^{\mu} \, \left( \, g_V
                        + g_A \gamma^5 \, \right) \, \Psi\, \Big( A_{(0)
                        \mu} \, + \, \sqrt{2} \, \sum_{n=1}^{\infty}
                        A_{(n) \mu} \Big) \, ,
\end{equation}
where  $A_{\mu}$  generically  denotes  some  gauge boson  and  g  the 
respective coupling constant.  The coupling parameters $g_V$ and $g_A$ 
are set by the electroweak quantum numbers of the fermions and receive 
their SM values. Because the KK mass  eigenmodes $\hat{A}_{(n)}^{\mu}$ 
generally differ  from  the   Fourier   modes    $A_{(n)}^{\mu}$    in  
(\ref{genericfermionint}), as we have  seen before, their couplings to
fermions $g_{V(n)}$ and $g_{A(n)}$ have to  be individually calculated 
for each model. The photon and its possible KK modes  are not affected 
by  spontaneous symmetry  breaking  and  keep their  simple couplings,
already presented in Fig. \ref{QEDvertices}. For the bulk-bulk  model,
the shifts in the vector and  axial-vector couplings of the  $Z$ boson
are actually the same, such that it is sufficient to replace the SU(2)
coupling constant $g$ by $g_{Z (n)}$ for each KK mode, in  analogy  to 
the Abelian Higgs model. The  mass generation  in the   Yukawa sector,
involving brane fermions, hardly changes at all.

An  even more minimal 5-dimensional  extension  of electroweak physics
constitutes a model in which only  the U(1)$_Y$ -sector feels the extra
dimension  while the  SU(2)$_L$ gauge  fields  are localized  at $y=0$
(brane-bulk model). It is described by the Lagrangian
\begin{equation}
\begin{split}
\label{electroweaklagrangianuoneinbulk}
\mathcal{L} (x,y) = - \frac{1}{4} & B_{M N} \, B^{M N} \! + \delta(y) \!
              \left[ - \frac{1}{4} \, F^a_{\mu \nu} F^{a \mu \nu} \!
	      + \! \left( D _{\mu} \Phi \right)^{\dagger} \!
	      \left( D^{\mu} \Phi \right) - V(\Phi) \right] \\[1ex]
&+ \, \mathcal{L}_{\mathrm{GF}}(x,y)\ +\ \mathcal{L}_{\mathrm{FP}}(x,y) \, ,
\end{split}
\end{equation}
where $\Phi$ denotes the Standard Model Higgs doublet on the brane and
the covariant derivative
\begin{equation}
D_{\mu} \, = \, \partial_{\mu}
\, - \, i \, g \,  A^a_{\mu}(x) \, \tau^a \, -  i \, \frac{g_5'}{2} \,
B_{\mu}(x,y)
\end{equation}
involves a  brane as well  as a bulk field.  The  $W$ bosons are brane
fields  and their  physics is  completely SM-like.  In this  case, the
Higgs field being charged with respect to both gauge  groups has to be
localized at   $y=0$ in order   to  preserve gauge  invariance  of the
(classical) Lagrangian.  A  gauge field on the brane cannot compensate 
the variation of a Higgs field under gauge transformations in the whole
bulk. For  the same reason, a bulk  Higgs is forbidden  in  the  third
possible  model  in   which  U(1)$_Y$  is  localized   while SU(2)$_L$ 
propagates in the fifth  dimension (bulk-brane model), i.e., the gauge
groups   interchange   their  r\^ole.   Consequently,  the  $W$ bosons
are bulk fields and are  described in analogy to the models  discussed
in  Sect. \ref{5DQED}.  In  both  models, after  spontaneous  symmetry 
breaking,  there  is  a  single  massless gauge field protected by the 
residual unbroken gauge symmetry, the  photon.  A second light neutral  
mode can be identified with the $Z$ boson. However, in contrast to the 
bulk-bulk  model,  there  is  only  a single neutral tower of heavy KK  
modes.  Up  to  small  admixtures  due  to  the  brane  VEV  (see Sect.  
\ref{5DQED}),  it  mainly  contains  the  U(1)$_{\mathrm{Y}}$  or  the
neutral SU(2)$_{\mathrm{L}}$ gauge field, respectively.  Nevertheless,
for simplicity,  we will  refer to it   as  $Z$-boson KK  tower. Note,
however, that  $g_{V(n)}$  and $g_{A(n)}$ in (\ref{genericfermionint})
are  affected differently for the  $Z$-boson and its   KK modes.  Most
easily they are parameterized by introducing effective quantum numbers
$T_{3(n)}$  and    $Q_{(n)}$  to absorb    all  the higher-dimensional
effects.  

After the setup of all the models, we can finally turn  our  attention 
to  the  actual  predictions   of   higher-dimensional   theories  for 
experiment.

\section{Effects on Electroweak Observables}
\label{Phenomenology}

\begin{figure}[t]
\begin{center}
\includegraphics[width=7.2cm]{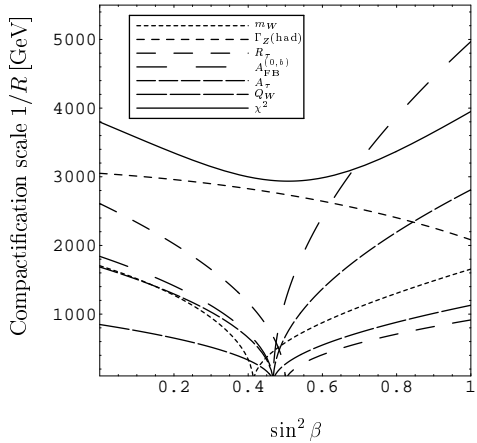}
\end{center}
\caption{\label{bulkbulkboundlot} Lower bounds on  $M=1/R$ (in TeV) from different 
observables at  the 3$\sigma$ confidence level for the bulk-bulk model }
\end{figure}

In this section, we will concentrate  on the phenomenology and present
bounds    on the   compactification scale   $M    =  1/R$  of  minimal
higher-dimensional extensions of   the SM, calculated  by  analyzing a
large number  of observables.  To be  specific, we proceed as follows.
We         relate              the            SM            prediction
$\mathcal{O}^{\mathrm{SM}}$~\cite{PDG,EWWG}  for an  observable to the
prediction  $\mathcal{O}^{\mathrm{HDSM}}$   for   the same  observable
obtained in the higher-dimensional SM under investigation through
\begin{equation}
\label{generalformofpredictions}
{\cal O}^{\rm HDSM} \ =\ {\cal O}^{\rm SM} \, \big( 1\: +\: 
\Delta^{\rm HDSM}_{\cal O} \big)\, .
\end{equation}
Here, $\Delta^{\rm HDSM}_{\cal O}$ is the tree-level modification of a
given observable ${\cal O}$ from its  SM value due  to the presence of
one  extra  dimension. The tree-level modifications can be expanded in
powers of the typical scale factor
\begin{equation}
\label{scalefactor}
X \, = \, \frac{\pi^2}{3} \frac{m^2_Z}{M^2} \, \, .  
\end{equation}
We work  to first  order in $X$  being  a very good  approximation for
phenomenologically  viable   compactification   scales   in the    TeV
region.   On the other   hand, to  enable a   direct comparison of our
predictions with precise data~\cite{PDG,EWWG}, we include SM radiative
corrections to~$\mathcal{O}^{\mathrm{SM}}$.    However, we neglect SM-
as well  as KK-loop contributions to  $\Delta^{\rm  HDSM}_{\cal O}$ as
higher order effects.

As  input  SM  parameters for our numerical predictions, we choose the 
most  accurately  measured  ones, namely the $Z$-boson mass $m_Z$, the 
electromagnetic  fine  structure  constant~$\alpha$,   and  the  Fermi 
constant $G_F$.  While $\alpha$  is  not affected  in the models under 
study,  $m_{Z}  = m_{Z (0)}$, the mass of the lightest mode in the $Z$ 
boson KK tower,   generally  deviates  from its SM form, where we have  
$m_Z^{\mathrm{SM}}  =  \sqrt{g^2 +  g'{}^2}\, v/2$ at  tree level.  To 
first order in $X$, $m_{Z}$ may be parameterized by
\begin{equation}
\label{deltazdef}
m_{Z} \  =\ m^{\mathrm{SM}}_{Z} \, \left( \, 1 \: + \: \Delta_{Z}\,X \, \right) \,, 
\end{equation}
where  $\Delta_{Z}$    is  a   model-dependent   parameter.   For  the  
bulk-bulk,  brane-bulk  and   bulk-brane   models  with respect to the 
SU(2)$_L$  and U(1)$_Y$  gauge groups, $\Delta_{Z}$  is given by
\begin{equation}
\Delta_{Z} = \left\{ \ - \, \frac{1}{2} \,  \sin^4\beta\,,\ - 
\, \frac{1}{2} \, \sin^2 \hat{\theta}_W\,,\ - \, \frac{1}{2} \, 
\cos^2\hat{\theta}_W\, \right\} \,,
\end{equation} 
where   $\hat{\theta}_W$  is an  effective weak   mixing   angle to be
introduced   below  in  (\ref{effmixinganle}).  These  shifts  in  the  
$Z$-boson  mass are induced by the VEV of a brane Higgs.

\begin{table}[t]
\caption{\label{branebulkboundplot} Lower bounds on  $M=1/R$ (in TeV) 
at the 3$\sigma$ confidence  level  for the brane-bulk and  bulk-brane
models}
\renewcommand{\arraystretch}{1.5}

\vspace{.2cm}
\begin{center}
\begin{tabular}{cccc} 
\hline
Observable & U(1)$_Y$ in bulk & \hspace{0.2cm} & SU(2)$_L$ in bulk \\ \hline 
$m_W$ & 1.2 & &  1.2\\ 
$\Gamma_Z (\mathrm{had})$ & 0.8 & & 2.3 \\ 
$Q_W (\mathrm{Cs})$ & 0.4 & & 0.8 \\ 
$A_{\mathrm{FB}}^{(0,b)}$ & 4.4 & & 2.4 \\ 
$A_{\tau}$ & 2.5 & & 1.4 \\ 
$R_{\tau}$ & 1.0 & & 0.5 \\ 
global analysis & 3.5 & & 2.6 \\ 
\hline
\end{tabular}
\end{center}
\end{table}

The  Fermi constant   $G_F$, as  determined by the  muon lifetime, may 
receive additional direct contributions due to KK states mediating the 
muon decay. We may  account for this  modification of $G_F$ by writing
\begin{equation}
G_F \ = \ G^{\mathrm{SM}}_F \, \left( \, 1\: +\: \Delta_G\, X \, \right) \,,
\end{equation}
where $\Delta_G$  is again model-dependent   and has to  be calculated
consistently to first order in $X$.

The relation  between the weak mixing angle   $\theta_W$ and the input
variables is also affected by the fifth dimension. Hence, it is useful
to define an effective mixing angle $\hat{\theta}_W$ by
\begin{equation}
\label{effmixinganle}
\sin^2 \hat{\theta}_W  \ =  \  \sin^2 \theta_W  \, \left( \,  1\: +\:
\Delta_{\theta}\,X \, \right) \, , 
\end{equation}
such that the effective angle still fulfills the tree-level relation
\begin{equation}
\label{definitionequationfonormalizedthetaw}
G_F \ = \ \frac{\pi \alpha}{\sqrt{2} \sin^2 \hat{\theta}_W \, \cos^2
\hat{\theta}_W \, m_{Z}^2} \ ,
\end{equation}
of the Standard Model.

\begin{table}[t]
\caption{\label{globalbounds}  Lower bounds (in TeV) on the
compactification scale $M = 1/R$ at 2$\sigma$, 3$\sigma$ and 5$\sigma$
confidence levels from combined precision observables}

\vspace{.2cm}
\renewcommand{\arraystretch}{1.5}
\begin{center}
\begin{tabular}{cccc} 
\hline
model & 2$\sigma$ & 3$\sigma$ & 5$\sigma$ \\ \hline 
SU(2)$_L$-brane, U(1)$_Y$-bulk & \hspace{0.2cm} 4.3
\hspace{0.2cm} & \hspace{0.2cm} 3.5 \hspace{0.2cm} & \hspace{0.2cm}
2.7 \hspace{0.2cm} \\ 
SU(2)$_L$-bulk, U(1)$_Y$-brane & 3.0 & 2.6 & 2.1 \\[-1ex]
\parbox{5cm}{\begin{center} SU(2)$_L$-bulk, 
U(1)$_Y$-bulk \\ (brane Higgs) \end{center}} & 4.7 & 4.0 & 3.1 \\[-3ex] 
\parbox{5cm}{\begin{center} SU(2)$_L$-bulk, 
U(1)$_Y$-bulk \\ (bulk Higgs) \end{center}} & 4.6 & 3.8 & 3.0 \\[-1ex] \hline
\end{tabular}
\end{center}
\end{table}

For  the tree-level  calculation of  $\Delta^{\rm  HDSM}_{\cal O}$, we
have  to  carefully consider the effects  from   mixing of the Fourier
modes on  the masses of the Standard-Model  gauge bosons as well as on
their couplings to  fermions.  In  addition, we have  to keep  in mind
that  the mass spectrum of the  KK  gauge bosons  also  depends on the
model under consideration.

Within the  framework outlined above,   we first compute  $\Delta^{\rm
HDSM}_{\cal O}$ for the following  high precision observables to first
order in $X$: the $W$-boson mass  $m_W$, the $Z$-boson invisible width
$\Gamma_Z  (\nu     \overline{\nu})$,     $Z$-boson  leptonic   widths
$\Gamma_Z(l^+l^-)$,          the  $Z$-boson     hadronic         width
$\Gamma_Z(\mathrm{had})$,  the weak charge  of cesium  $Q_W$ measuring
atomic parity  violation, various  ratios  $R_l$ and  $R_q$  involving
partial $Z$-boson widths, fermionic asymmetries $A_f$ at the $Z$ pole,
and    various        fermionic       forward-backward     asymmetries
$A_{\mathrm{FB}}^{(0,f)}$.  For  example, for the invisible  $Z$ width
$\Gamma_Z(\nu \overline{\nu})$ we obtain
\begin{equation}
\Delta^{\rm HDSM}_{\Gamma_Z(\nu \overline{\nu})} = \left\{
\begin{array}{cl}
\sin^2 \! \hat{\theta}_W \, \left( \, \! \sin^2 \! \beta - 1 \! \, \right)^2 
- 1 & \mbox{for  the bulk-bulk  model,}\\[1ex]
- \sin^2 \! \hat{\theta}_W & \mbox{for the  brane-bulk model,} \\[1ex]
- \cos^2 \! \hat{\theta}_W & \mbox{for the bulk-brane model.}
\end{array}
\right.
\end{equation}

Employing the   results for $\Delta^{\mathrm{HDSM}}_{\mathcal{O}}$ and
calculating all the electroweak observables considered in our analysis
by   virtue  of~(\ref{generalformofpredictions}), we  confront   these
predictions with the  respective  experimental results. We  can either
test  each variable individually or perform  a $\chi^2$ test to obtain
combined bounds on the compactification scale $M=1/R$, where
\begin{equation}
  \label{chi2}
\chi^2(R) \ = \ \sum_i \,\frac{\left( \, \mathcal{O}_i^{\mathrm{exp}}\: -\:
\mathcal{O}_i^{\mathrm{HDSM}} \, \right)^2}
{\left( \, \Delta \mathcal{O}_i \, \right)^2} \, ,
\end{equation}
$i$  runs over all the observables,  and $\Delta \mathcal{O}_i$ is the
combined experimental and theoretical error. A compactification radius
is considered to be compatible at the $n \sigma$ confidence level (CL)
if    $\chi^2     (R)   -   \chi^2_{\mathrm{min}}    <    n^2$,  where
$\chi^2_{\mathrm{min}}$   is   the    minimum    of  $\chi^2$  for   a
compactification radius in the physical region, i.e.~for $R^2>0$.

Figure~\ref{bulkbulkboundlot}  summarizes  the   lower  bounds  on the
compactification scale $M$ inferred from different types of observables 
for the  bulk-bulk model.  In this  model,  we present the bounds as a 
function   of   $\sin^2\beta$   parameterizing   the  Higgs sector. In 
Table~\ref{branebulkboundplot},  we summarize  the  bounds obtained by
our calculations for the two   bulk-brane models. The bounds from  the
global analysis at      different confidence  levels are    shown   in
Table~\ref{globalbounds}. The global $2 \sigma$  bounds lie in the  $4
\sim 5$~TeV region,   only the bulk-brane  model   is less restricted.
Here,    a compactification scale   of   3~TeV cannot  be excluded  by
electroweak precision observables.

\begin{table}[t]
\caption{\label{fermiondata} 2$\sigma$ bounds in TeV  inferred 
from fermion-pair production}

\vspace{.2cm}
\begin{center}
\begin{tabular}{ccccc} 
\hline
\parbox{5cm}{\begin{center} \vspace{-0.05cm} model \vspace{-0.05cm} \end{center}}
& $\mu^+ \mu^-$ & $\tau^+ \tau^-$ & hadrons & $e^+ e^-$ \\ \hline
\parbox{5cm}{\begin{center} SU(2)$_L$-brane, U(1)$_Y$-bulk \end{center}} & 
\hspace{0.2cm} 2.0 \hspace{0.2cm} & \hspace{0.2cm} 2.0 \hspace{0.2cm} & 
\hspace{0.2cm} 2.6 \hspace{0.2cm} & \hspace{0.2cm} 3.0 \hspace{0.2cm} \\[-2ex]
\parbox{5cm}{\begin{center} SU(2)$_L$-bulk, U(1)$_Y$-brane  \end{center}} & 
1.5 & 1.5 & 4.7 & 2.0 \\[-2ex] 
\parbox{5cm}{\begin{center} SU(2)$_L$-bulk, U(1)$_Y$-bulk \\ (brane Higgs) 
 \end{center}} & 
2.5 & 2.5 & 5.4 & 3.6 \\[-3ex]
\parbox{5cm}{\begin{center} SU(2)$_L$-bulk, U(1)$_Y$-bulk \\ (bulk Higgs)  
\end{center}} & 
2.5 & 2.5 & 5.8 & 3.5 \\[-1ex] \hline
\end{tabular}
\end{center}
\end{table}

While the precision observables, analyzed so  far, are measured at the
$Z$ pole or even at low  energies, LEP2 provides us with data on cross 
sections  at   higher  energies,  up  to more   than 200 GeV. At these 
energies,  the  interference  terms  between  Standard  Model  and  KK
contributions to a process  like fermion-pair production  dominate the
higher-dimensional   effects.  In a    first  approximation, they  are 
only suppressed by a factor of order $s/M^2$, where $\sqrt{s}$ is  the  
center-of-mass   energy,   compared  to  the   typical  scale   factor 
$X$  of mass mixings and coupling shifts.  Higher  energies  naturally 
lead to  more  sensitivity with  respect to a possible fifth dimension
\cite{CL,MPR2,MN}.  As   the  simplest example   for  LEP2  processes, 
let us have a closer look  at fermion-pair  production.  The  relevant 
differential  cross  section  for  these  observables is given at tree 
level by
\begin{table}[t]
\caption{\label{combineddata} 2$\sigma$ bounds in TeV  inferred 
from precision observables and LEP2 cross sections}

\vspace{.2cm}
\renewcommand{\arraystretch}{1.5}
\begin{center}
\begin{tabular}{cccc} 
\hline
model & LEP1 & LEP2 & combined \\ \hline 
SU(2)$_L$-brane, U(1)$_Y$-bulk & \hspace{0.2cm} 4.3
\hspace{0.2cm} & \hspace{0.2cm} 3.5 \hspace{0.2cm} & \hspace{0.2cm}
4.7 \hspace{0.2cm} \\ 
SU(2)$_L$-bulk, U(1)$_Y$-brane & 3.0 & 4.4 & 4.3 \\[-1ex]
\parbox{5cm}{\begin{center} SU(2)$_L$-bulk, 
U(1)$_Y$-bulk \\ (brane Higgs) \end{center}} & 4.7 & 5.4 & 6.1 \\[-3ex] 
\parbox{5cm}{\begin{center} SU(2)$_L$-bulk, 
U(1)$_Y$-bulk \\ (bulk Higgs) \end{center}} & 4.6 & 5.7 & 6.4 \\[-1ex] \hline
\end{tabular}
\end{center}
\end{table}
\begin{equation}
\label{diffcrosssection}
\begin{split}
\frac{\sigma (e^+ e^- \to f \, \overline{f})}{d \, \cos \theta} = 
\frac{s}{128 \, \pi} 
& \left[( 1 + \cos \theta)^2 \, (|M^{ef}_{LL} (s)|^2 + |M^{ef}_{RR} (s)|^2) \, \, + 
\right. \\
& \, \, \, \left. ( 1 - \cos \theta)^2 \, (|M^{ef}_{LR} (s)|^2 + |M^{ef}_{RL} (s)|^2) 
\right] \, ,
\end{split}
\end{equation}
where  $\theta$  is the scattering angle between the incoming electron
and  the negatively  charged outgoing fermion, and
\begin{equation}
\label{effprop}
M^{ef}_{\alpha \beta} (s) =  
\sum_{n=0}^{\infty}  \left( \, e_{(n)}^2 \, \frac{Q_e Q_f}{s - m^2_{\gamma (n)}} \, + \, 
\frac{g^e_{\alpha (n)} g^f_{\beta (n)}}{\cos^2 \theta_W} \, 
\frac{1}{s - m^2_{Z (n)}} \, \right) \, .
\end{equation}
The couplings $g^f_{L (n)}$ and $g^f_{R (n)}$ in turn are given by
\begin{equation}
\label{couplings}
\begin{split}
g^f_{L (n)} & = g_{Z (n)} \left( \, T_{3f (n)} - Q_{f (n)} \sin^2 \theta_W \, 
\right) \, , \\
g^f_{R (n)} & = g_{Z (n)} \left( \, - Q_{f (n)} \sin^2 \theta_W \, \right) \, .
\end{split}
\end{equation}
$T_{3f (n)}$,  $Q_{f (n)}$, $g_{Z   (n)}$, $e_{(n)}$, $m_{Z  (n)}$ and
$m_{\gamma (n)}$, as introduced in Sect. \ref{MinimalExtensions},  can
be  calculated  to  first  order in $X$, e.g.  from the exact analytic
expressions in \cite{MPR}.    For Bhabha  scattering  the  $t$-channel
exchange  also has  to  be taken into account,   however, there are no
fundamental differences. The   above  parameterization for  the  cross
section is particularly convenient  because  it clearly separates  the
higher-dimensional   effects   from  well   known  physics.   All  the  
higher-dimensional physics  manifests  itself  in the effective sum of 
s-channel  propagators (\ref{effprop}) differing  with  respect to the 
Standard Model.

From (\ref{diffcrosssection}), we calculate $\Delta^{\rm HDSM}_{\cal O}$
for  the  different  fermion-pair production channels at LEP2 energies 
and  finally  the  bounds  shown  in Table \ref{fermiondata}. In Table 
\ref{combineddata},  all   the   bounds  are combined  for final lower  
limits on  the  compactification scale. The  bounds for the brane-bulk 
and  the  bulk-brane models  from  LEP1 and LEP2  observables are kind 
of complementary, such that the combined limit is larger than 4~TeV in  
both models.  The  compactification  scale for the bulk-bulk model, no  
matter where the Higgs  lives, is still  more restricted to lie  above 
6~TeV at the $2 \sigma$ confidence level.

\section{Conclusions}
\label{Conclusions}

The aim of the  present notes has been  to give an introduction to the
model-building of  low-energy   5-dimensional electroweak  models.  We
have derived  step by step  the corresponding  four dimensional theory
by compactifying on the  orbifold $S^1/Z_2$ and by integrating out the
extra dimension.   We have   paid  special attention   to consistently
quantize the  higher-dimensional  models in the  generalized  $R_\xi$
gauges.  The 5-dimensional  $R_\xi$ gauge fixing conditions introduced
here  lead, after compactification, to  a  4-dimensional Lagrangian in
the standard  $R_\xi$    gauge for  each KK mode.     The  latter also
clarifies the r\^ole of the different degrees  of freedom.  One of the
main advantages of  our gauge-fixing  procedure is  that one  can  now
derive manifest gauge-independent analytic expressions for the KK-mass
spectrum  of  the  gauge bosons  and   for their  interactions to  the
fermionic matter. Most importantly,  one  may even apply an  analogous
gauge-fixing approach to spontaneous symmetry breaking theories.

To render the topics under discussion more intuitive, we have analyzed
all  the main  ideas  in simple Abelian  toy  models. However, we have
pointed     out how   to  generalize  these    ideas   to new possible
5-dimensional extensions of the SM in which the SU(2)$_L$ and U(1)$_Y$
gauge  fields and Higgs   bosons  may or may  not  all  experience the
presence of  the fifth dimension.  The fermions  in all the models are
considered to  be confined  to  one  of   the two boundaries   of  the
$S^1/Z_2$ orbifold.

After introducing a  framework  for deriving predictions  of  possible
observables, we  have  given   a glimpse of   higher-di\-men\-sio\-nal
phenomenology. Electro\-weak  precision observables are  considered as
well  as cross  sections  for fermion-pair  production   at LEP2.   In
particular, we have presented bounds  on the compactification scale $M
= 1/R$ in three different  5-dimensional extensions of the SM: (i)~the
SU(2)$_L\otimes$U(1)$_Y$-bulk model, where  all  SM gauge  bosons  are
bulk fields; (ii) the SU(2)$_L$-brane, U(1)$_Y$-bulk model, where only
the     $W$ bosons are restricted to     the  brane, and (iii)~the
SU(2)$_L$-bulk,  U(1)$_Y$-brane model, where  only the  U(1)$_Y$ gauge
field is  confined to the brane.  For the often-discussed first model,
we find the  2$\sigma$ lower  bounds on $M$:   $M\stackrel{>}{{}_\sim}
6.4$~and 6.1~TeV, for a  Higgs  boson living in the   bulk and on  the
brane,  respectively.  For    the    second and third   models,    the
corresponding 2$\sigma$ lower  limits are 4.7~and 4.3~TeV. Hence,  the
bounds for different models can differ significantly.

Any   non-stringy   field-theoretic  treatment of   higher-dimensional
theories, as the one presented here, involves a number of assumptions.
Although the results   obtained in the  higher-dimensional models with
one compact dimension are  convergent at the  tree level,  they become
divergent if more than one extra dimensions are considered.  Also, the
analytic results are ultra-violet (UV) divergent at the quantum level,
since the higher-dimensional  theories are not renormalizable.  Within
a string-theoretic framework, the above UV divergences are expected to
be regularized by  the string  mass  scale $M_s$.  Therefore, from  an
effective field-theory point of view, the phenomenological predictions
will depend  to some  extend  on the UV  cut-off  procedure~\cite{KMZ}
related to the string scale $M_s$.  Nevertheless, assuming validity of
perturbation  theory, we expect that quantum  corrections due to extra
dimensions will not exceed the 10\% level of the tree-level effects we
have   been   studying  here.    Finally,   we  have  ignored possible
model-dependent  winding-number contributions~\cite{ABL} and radiative
brane effects~\cite{Georgi} that might also affect  to some degree our
phenomenological predictions.

The lower limits on the compactification scale  derived by the present
global  analysis indicate that  resonant  production of  the first  KK
state may be at the edge  of accessibility at the  LHC, at which heavy
KK masses up  to 6--7~TeV~\cite{AB,RW} might  be explored.  Hence, the
phenomenological analysis  has to be  carried  further in order to  be
able to  discriminate possible  higher-dimensional signals from  other
Standard Model extensions.

\subsection*{Acknowledgements}
This  work was supported  by the  Bundesministerium f\"ur  Bildung and
Forschung (BMBF,  Bonn, Germany) under the  contract number
05HT1WWA2.


\begin{thebibliography}{8.}
\addcontentsline{toc}{section}{References}

\bibitem{KK} T. Kaluza: Sitzungsber. d. Preuss. Akad. d. Wiss. Berlin, 
966 (1921) O. Klein: Zeitschrift f. Physik \textbf{37} 895 (1926)  

\bibitem{review} For a review, see e.g., M.B. Green, J.H. Schwarz,
  E. Witten: \emph{Superstring Theory}. (Cambridge University
  Press, Cambridge 1987).

\bibitem{IA} I. Antoniadis: Phys. Lett. B \textbf{246}, 377 (1990)

\bibitem{JL} J.D. Lykken: Phys. Rev. D \textbf{54} 3693 (1996)
  
\bibitem{EW} E. Witten: Nucl. Phys. B \textbf{471} 135 (1996) 
  P. Ho$\check{{\rm r}}$ava, E. Witten: Nucl. Phys. B \textbf{460}
  506 (1996); Nucl. Phys. B \textbf{475} 94 (1996)

\bibitem{ADS} N. Arkani-Hamed, S. Dimopoulos, G. Dvali: Phys. Lett. 
  B \textbf{429} 263 (1998) I. Antoniadis, N. Arkani-Hamed, S. Dimopoulos,
  G. Dvali: Phys. Lett. B \textbf{436} 257 (1998); N. Arkani-Hamed,
  S. Dimopoulos, G. Dvali: Phys. Rev. D \textbf{59} 086004 (1999)

\bibitem{DDG} K.R. Dienes, E.  Dudas, T. Gherghetta: Phys. Lett.
  B \textbf{436} 55 (1998); Nucl. Phys. B \textbf{537} 47 (1999)
    
\bibitem{AB} I. Antoniadis, K. Benakli: Int. J. Mod. Phys.
    A \textbf{15} 4237 (2000)
  
\bibitem{AKT} I. Antoniadis, E. Kiritsis, T.N. To\-ma\-ras: 
  Phys. Lett. B \textbf{486} 186 (2000)

\bibitem{NY} P. Nath, M. Yamaguchi: Phys. Rev. D \textbf{60} 116006 
  (1999); Phys. Lett. B \text{466} 100 (1999)
  
\bibitem{WJM} W.J. Marciano: Phys. Rev. D \textbf{60} 093006 (1999); M.
  Masip, A. Pomarol: Phys. Rev. D \textbf{60} 096005 (1999)
  
\bibitem{CCDG} R. Casalbuoni, S. De Curtis, D. Dominici, R. Gatto:
  Phys. Lett. B \textbf{462} 48 (1999); C. Carone: Phys. Rev.
  D \textbf{61} 015008 (2000)

\bibitem{DPQ2} A. Delgado, A. Pomarol, M. Quiros: JHEP \textbf{0001}
  030 (2000)
    
\bibitem{RW} T. Rizzo, J. Wells: Phys. Rev. D \textbf{61} 016007 (2000); 
  A. Strumia: Phys. Lett. B \textbf{466} 107 (1999)
  
\bibitem{DPQ1} A. Delgado, A. Pomarol, M. Quiros: Phys. Rev. D 
  \textbf{60} 095008 (1999)
  
\bibitem{CL} K. Cheung, G. Landsberg: 
  Phys. Rev. D \textbf{65} 076003 (2002)

\bibitem{MPR} A. M\"uck, A. Pilaftsis, R. R\"uckl:
  Phys. Rev. D \textbf{65} 085037 (2002)

\bibitem{GGH} For example, see H. Georgi, A.K. Grant, G. Hailu:
  Phys. Lett. B \textbf{506} 207 (2001)

\bibitem{GNN} D.M. Ghilencea, S. Groot Nibbelink, H.P. Nilles:
  Nucl. Phys. B \textbf{619} 385 (2001)

\bibitem{PS} J. Papavassiliou, A. Santamaria:
  Phys. Rev. D \textbf{63} 125014 (2001)

\bibitem{DMN} D. Dicus, C. McMullen, S. Nandi: 
  Phys. Rev. D \textbf{65} 076007 (2002)

\bibitem{ACD} T. Appelquist, H.C. Cheng and B.A. Dobrescu:
  Phys. Rev. D \textbf{64} 035002 (2001)

\bibitem{DCH} R.S. Chivukula, D.A. Dicus, H.-J. He:
  Phys. Lett. B \textbf{525} 175 (2002)
  
\bibitem{PDG} Particle Data Group (D.E. Groom et al.): European
  Physical Journal C \textbf{15} 1 (2000)

\bibitem{EWWG} The LEP Collaborations ALEPH, DELPHI, L3, OPAL, the LEP
  Electroweak Working Group and the SLD Heavy Flavor and Electroweak
  Groups: hep-ex/0112021

\bibitem{MPR2} A. M\"uck, A. Pilaftsis, R. R\"uckl:
  work in preparation

\bibitem{MN} C.D. McMullen, S. Nandi: hep-ph/0110275

\bibitem{KMZ} T. Kobayashi, J.  Kubo, M.  Mondragon and G.  Zoupanos,
  Nucl.\ Phys.\ B {\bf 550} 99 (1999)

\bibitem{ABL} I. Antoniadis, K. Benakli and A. Laugier,
  JHEP {\bf 0105} 044 (2001) 

\bibitem{Georgi} H. Georgi, A.K. Grant, G. Hailu,
  Phys.\ Lett.\ B {\bf 506} 207 (2001)
  M. Carena, T. Tait, C.E.M. Wagner: hep-ph/0207056


\end{thebibliography}
\end{document}